\documentstyle[12pt,epsf]{article}

\advance\voffset by -1.5cm
\advance\hoffset by -1.5cm
\def\be{\begin{equation}}
\def\ee{\end{equation}}

\def\pmb#1{\setbox0=\hbox{#1}
 \kern-.025em\copy0\kern-\wd0
 \kern.05em\copy0\kern-\wd0
 \kern-.025em\raise.0433em\box0 }

\def\3{\ss}
\def\sq{\hbox{\rlap{$\sqcap$}$\sqcup$}}
\def\qed{\ifmmode\sq\else{\unskip\nobreak\hfil
\penalty50\hskip1em\null\nobreak\hfil\sq
\parfillskip=0pt\finalhyphendemerits=0\endgraf}\fi}
\def\half {\frac{1}{2}}

\def\bbbz {{\sf Z\!\!Z}}

\def\bbbr {{\rm I\!R}}

\def\ss{\bf S}

\def\C{{\cal C}}
\def\M{{\cal M}}

\newtheorem{thm}{Theorem}

\begin{document}

\thispagestyle{empty}
\def\thefootnote{\fnsymbol{footnote}}
\begin{flushright}
  hep-th/9802033\\
  HUTP-98/A007
 \end{flushright}
\vskip 0.5cm

\begin{center}\LARGE
{\bf String Junctions and BPS States in 
Seiberg-Witten Theory}
\end{center}
\vskip 1.0cm
\begin{center}
{\large  Oren Bergman\footnote{E-mail  address: {\tt
bergman@string.harvard.edu}} and Ansar Fayyazuddin
\footnote{E-mail  address: 
{\tt ansar@curie.harvard.edu}}}

\vskip 0.5 cm
{\it Lyman Laboratory of Physics\\
Harvard University\\
Cambridge, MA 02138}
\end{center}

\vskip 1.5cm

\begin{center}
February 1998
\end{center}

\vskip 1.5cm

\begin{abstract}
We argue that certain BPS states in the D3-brane 
probe realization of N=2 SU(2) Super-Yang-Mills 
theory correspond to multi-pronged strings 
connecting the D3-brane to the background 7-branes.
This provides a physical realization of the decay of 
these states on the curve of marginal stability,
and explains their absence in the strong coupling 
regime. 
\end{abstract}

\vskip 1.5cm 
\begin{center}
 PACS codes: 11.25.-w, 11.15.-q, 11.30.Pb
\end{center}

\vfill
\setcounter{footnote}{0}
\def\thefootnote{\arabic{footnote}}
\newpage

\renewcommand{\theequation}{\thesection.\arabic
{equation}}

\section{Introduction}
\setcounter{equation}{0}

\noindent String theory provides new ways of 
approaching problems 
in field theory which are hard and often 
intractable using 
traditional field theory techniques.
A case in point is the BPS spectrum
of four dimensional $N=4$ Super-Yang-Mills theory,
with gauge group $SU(n)$. The conjectured 
$SL(2,\bbbz)$
symmetry of this theory predicts the existence of 
a host
of 1/2 BPS dyon states, {\it i.e.} states 
preserving 1/2 of
the total supersymmetry. The existence of these 
states,
however,
has only been rigorously demonstrated in the one and 
two monopole cases \cite{sendyon}, 
although there are indications 
that
that they exist as well in the multi-monopole case
\cite{multimonopole,ferrari}.
In string theory one can study $N=4$ SYM with
gauge group $SU(n)$ as the low-energy world-volume
theory of $n$ parallel D3-branes \cite{tseytlin,gg}. 
The above BPS states
correspond simply to type IIB strings between pairs 
of 
D3-branes. Since type IIB strings can carry 
arbitrary
$(p,q)$ charges, where $p$ and $q$ are relatively 
prime
\cite{wittenbound},
it follows that the 1/2 BPS states in the field 
theory
carry such charges as well, and this gives the 
complete
spectrum of 1/2 BPS states.

For $n>2$, the $N=4$ superalgebra \cite{osborne}
also allows states 
preserving only 1/4 of the supersymmetry, {\it i.e.} 
1/4 BPS 
states \cite{fh}.
One characteristic of these states is that they carry
mutually non-local charges under the different $U(1)$
subgroups of $SU(n)$. It would therefore be difficult
to establish their existence  from the field theory 
point of view.
On the other hand type IIB string theory contains 
objects
carrying mutually non-local $(p,q)$ charges; these
are the
so-called multi-pronged strings 
\cite{asy,schwarz}. 
In \cite{bergman}
it was shown that three-pronged strings ending on
three parallel D3-branes indeed correspond 
to 1/4 BPS states in the world-volume $N=4$
$SU(3)$ SYM theory.
This result can be generalized to $SU(n)$ and 
multi-pronged
strings, with up to $n$ prongs \cite{bergman2}.

In this paper we address the BPS spectrum of 
Seiberg-Witten theory \cite{sw}, 
namely $N=2$ SYM with
gauge group $SU(2)$, in the same approach.
In field theory the spectrum can be computed
in the weak coupling regime, but becomes 
inaccessible when the coupling is strong.  
In fact it has been argued that the strong
coupling BPS spectrum is different from the
weak coupling one, and that the spectrum
``jumps'' as a curve of marginal stability is crossed
\cite{sw,af,bilalf}.
The classical moduli space of the theory is 
parameterized by
a complex number $z$, and has a 
singularity at the origin corresponding to a point
of enhanced gauge symmetry. Quantum mechanically,
however, this singularity is split into two points,
located at $z=\pm \Lambda^2$, where
hypermultiplets carrying charges
$(0,1)$ and $(2,1)$ become massless. 
The exact low-energy
effective action is expressed in terms of an analytic
function $a(z)$ and its magnetic dual $a_D(z)$,
which can be expressed as integrals of a meromorphic 
differential $\lambda$ over cycles of  an auxiliary 
Riemann surface.
%
%\begin{equation}
%a(z) = \int_{\alpha}\lambda \quad , \quad
%   a_D(z) = \int_{\beta}\lambda \; .
%\end{equation} 
%
In particular the SYM coupling constant is given by
$\tau = da_D/da$. These functions also determine
the mass of a BPS state with electric charge $p$ and 
magnetic charge $q$, as
\begin{equation}
m_{(p,q)}=| pa(z) + qa_{D}(z)| \; .
\label{swmass}
\end{equation}
The spectrum in the weakly coupled regime
consists of electrically charged W-boson vector 
multiplets \footnote{We will work in the convention
in which the charge of the W-boson is $2$.},
and hypermultiplets with magnetic charge $\pm 1$ and 
electric charge $2n$ with $n\in \bbbz$.  
The moduli space is divided into two regions
by a curve of marginal stability $\C$, given by
$\mbox{Im}(a_D/a)=0$, which is
diffeomorphic to a circle.
It follows from (\ref{swmass}) and the triangle
inequality that the kinematic threshold for
the decay of BPS states into lighter BPS states
is saturated on this curve.
It is also apparent that since 
$pa+qa_D=0$ at the location of a $(p,q)$
singularity, this curve must pass through both
hypermultiplet points.
On this curve $a_D/a$ takes values between
$-2$ and $2$ \cite{af}. 
Specifically, on the upper half plane
part of $\C$, $a_D/a\in [-2,0]$, and on the lower half
plane part, $a_D/a\in [0,2]$.
Let us denote the region outside $\C$, {\it i.e.}
the weak coupling region, by $\M_+$, and the region 
inside $\C$, {\it i.e.} the strong coupling region, 
by $\M_-$. 
The BPS spectrum is expected to be different
in the strong coupling regime \cite{sw}. In fact it
was conjectured in \cite{af}, and shown in \cite{bilalf}, 
that the
spectrum in $\M_-$ consists solely of the $(0,1)$
and $(2,1)$
hypermultiplets,
which become massless at $z=\pm\Lambda^2$. 
As a consequence, all other BPS states which exist
in $\M_+$ must decay on the curve of marginal 
stability.

In string theory, $N=2$ $SU(2)$ SYM can be
understood as the low-energy theory on a 
D3-brane probe in the background of an orientifold
7-plane ($\Omega$7) \cite{bds}
\footnote{For an alternative approach to 
Seiberg-Witten theory and its BPS spectrum 
using six-dimensional 
self-dual strings see \cite{selfdual},
and for an approach using M-theory see 
\cite{wittenm}.}.
Sen has shown that this
classical background is modified quantum
mechanically, in accordance with F-theory,
to a background of two mutually non-local 7-branes
\cite{senf}.
In the world-volume theory of the D3-brane this
corresponds precisely to the singularity splitting of
Seiberg-Witten theory. The two hypermultiplets
which become massless at the two singularities
correspond to open strings beginning on the D3-brane
and ending on the respective 7-brane \cite{senbps}.
Since there are now three branes, one could in
principle construct states from multi-pronged 
strings 
connecting them. 
We shall demonstrate that, other than the above two
hypermultiplets, all the BPS states, including the 
W-bosons,
correspond to multi-pronged strings connecting
the D3-brane with the two 7-branes.
Furthermore, this approach provides a physical
realization of the decay of these BPS states
on the curve of marginal stability.

The paper is organized as follows.
In section~2 we review the D3-brane probe approach
and the picture of BPS states as open strings.
In section~3 we discuss multi-pronged strings
in 7-brane backgrounds, and derive the supersymmetry 
conditions. In section~4 we show that the W-boson
corresponds to a certain four-pronged string, and in
section~5 we similarly construct multi-pronged
string representations for the massive hypermultiplet
dyons.
Section~6 contains our summary and conclusions.

\vspace{0.3cm}

\noindent Note added: We were informed that results
similar to ours  will be reported in \cite{sethi}.

\section{Review of 3-brane probe}
\setcounter{equation}{0}

\noindent Consider a D3-brane probe in the background
of an $\Omega$7-plane in type IIB string theory.
Eight of the 32 supersymmetries are preserved, and
the theory on the world-volume of the D3-brane 
is $N=2$ $SU(2)$ SYM.
The Yang-Mills coupling and theta angle are related
to the string coupling and RR scalar as
\be
 \tau_{YM} = \tau_s\; ,
\ee
where 
$\tau_{YM}\equiv  \theta_{YM}/2\pi + i4\pi/g_{YM}^2$,
and $\tau_s \equiv  A + i/g_s$.
The classical moduli space is given by the position 
of the D3-brane relative to the $\Omega$7-plane, 
which we parameterize by the complex coordinate $z$.
At a generic point the gauge group is broken
to $U(1)$, and when the brane coincides with the 
plane ($z=0$) it is enhanced to $SU(2)$. 
The photon corresponds to an open string 
beginning and ending on the D3-brane, and the W-bosons 
correspond to a string, of either
orientation, which begins on the D3-brane,
reflects off the $\Omega$7-plane, and ends
on the D3-brane with the same orientation.
One can also introduce D7-branes into the background.
These will give rise to hypermultiplets (quarks)
in the D3-brane world-volume, corresponding to
open strings beginning on the D3-brane and ending
on the D7-branes (see fig.~1).
\begin{figure}[htb]
\epsfysize=4.0cm
\centerline{\epsffile{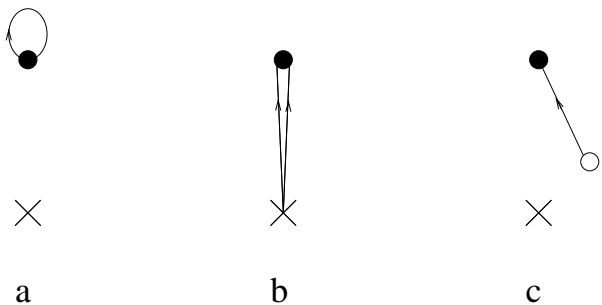}}
\caption{D3-brane (black circle) in the background
of an $\Omega$7-plane (cross) and a D7-brane 
(white circle). The BPS states are (a) photon,
(b) W-boson, and (c) quark.}
\end{figure}

In such a background the string coupling
acquires a $z$ dependence, which near
the $\Omega$7-plane or a D7-brane is given by
\be
 \tau(z) \sim {\alpha\over 2\pi i}
   \ln(z-z_i) \; ,
\ee
where $\alpha=1$ for a D7-brane and 
$\alpha=-4$ for the $\Omega$7-plane.
If four D7-branes coincide with the 
$\Omega$7-plane, the RR charge is canceled
locally, and $\tau$ is everywhere constant.
However, as pointed out by Sen \cite{senf}, 
if one tries to move the D7-branes away
from the orientifold plane, then 
sufficiently
near the $\Omega$7-plane $\mbox{Im}(\tau)$
becomes negative. The classical picture must
therefore be modified 
non-perturbatively.

The appropriate modification is specified by 
F-theory. Consider a compactification of F-theory
on a K3, which is given by a torus fibered over
$P^1$. This is equivalent to a 
compactification of type IIB string theory on
$P^1$ with 24 7-branes at positions $z_i$,
where the fiber degenerates. In the 
$T^4/\bbbz_2$ orbifold limit of K3 the
7-branes divide into four groups of six coinciding
7-branes. This limit is equivalent to a type
IIB orientifold compactification on 
$T^2/\bbbz_2$, where the $\bbbz_2$ action
is given by $(-1)^{F_L}\Omega R$. In this language
the four fixed points are $\Omega$7-planes,
which carry $-4$ units of RR charge. Consistency
requires the addition of four D7-branes at each 
fixed point, in order to cancel the RR charge.
We end up with precisely (four times) the classical
background considered above. It is then clear from 
the F-theory point of view how to deform this 
configuration; one simply moves away from the
orbifold limit of the K3 into a smooth K3.
The equation governing such deformations
close to one of the fixed points is given by
\be
 y^{2} = x^3 + f(z)x + g(z) \; ,
\label{deform}
\end{equation}
where $f$ and $g$ are polynomials of degree $2$ 
and $3$ respectively.  
The number of
deformations is given by the number of coefficients
in the polynomials $f$ and $g$, 
modulo shifting  $z$, and rescaling $z,x$, and $y$.
This gives five complex parameters, one of which
is the asymptotic value of $\tau$. The remaining
four parameters specify the
positions of the six 7-branes.  Thus we have fewer 
parameters than 7-branes.

It was noted in \cite{senf} that eq.~(\ref{deform})
is identical to the equation
for the Seiberg-Witten curve for $SU(2)$ with
$N_f=4$. Furthermore one can identify the four
parameters with the masses of the four quarks.
In analogy with the Seiberg-Witten solution,
one concludes therefore that the classical
gauge enhancement singularity, given in this 
case by the $\Omega$7-plane, splits into two
mutually non-local hypermultiplet points,
given in this case by two 7-branes with 
different $(p,q)$ charges. These can be taken
to be $(0,1)$ and $(2,1)$, in the convention
that a D7-brane has charges $(1,0)$.
The appearance of the Seiberg-Witten curve is
not a coincidence, and follows from the fact that
the theory on the D3-brane probe in this background
is precisely Seiberg-Witten theory \cite{bds}.
\begin{figure}[htb]
\epsfysize=4.5cm
\centerline{\epsffile{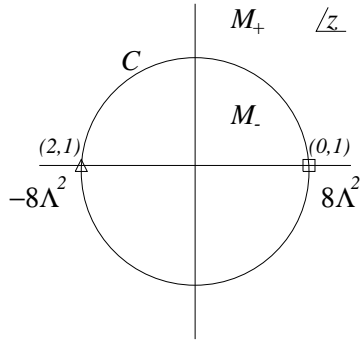}}
\caption{Quantum-mechanical resolution of the 
$\Omega$7-plane into a $(0,1)$ 7-brane (square)
at $z=8\Lambda^2$, and a $(2,1)$ 7-brane (triangle)
at $z=-8\Lambda^2$. The curve of marginal stability
$\C$ separates the weak coupling region $\M_+$
from the strong coupling region $\M_-$.}
\end{figure}

A picture of BPS states in the probe theory was 
developed in
\cite{senbps}.  There it was argued that a $(p,q)$ 
hypermultiplet dyon state 
in the probe theory corresponds to a $(p,q)$ string 
stretched between the D3-brane
probe and a 7-brane.  
Note that $(p,q)$ need not coincide with the
charges of the 7-brane, as the string can undergo
a monodromy when encircling a 7-brane.
A $(p,q)$ state exists as long as there exists
a path along which the total monodromy transforms
the $(p,q)$ charges of the string to the charges 
of the 7-brane. For this state to be BPS 
saturated the path should correspond to a $(p,q)$
``geodesic''. In other words it should minimize
the mass of the $(p,q)$ string, given by
\begin{equation}
m_{(p,q)}= \int_P T_{(p,q)}ds \;,
\end{equation}  
where $T_{(p,q)}$ is the tension of a $(p,q)$
string
\begin{equation}
T_{(p,q)} = \frac{1}{\sqrt{\mbox{Im}(\tau)}}
    |p+q\tau| \; .
\end{equation}
It is understood that the $(p,q)$
charges of the string
may change along the path $P$.
The metric $ds^2$ is given by \cite{greene}
\begin{equation}
 ds^2 = \mbox{Im}(\tau)\left|
 \frac{\eta^2 (\tau)}{\sqrt{2}\eta^2 
 (\tau_0)}\prod_{i=1}^{6}
 (z-z_i)^{-1/12}dz\right|^2 \; ,
\end{equation}
where $z_i$ are the locations of the six 7-branes, 
and $\tau_0$ is
the asymptotic value of $\tau$. We are primarily 
interested in the case $N_f=0$, which corresponds
to removing all four D7-branes to infinity. As shown in
\cite{ansar}, the metric in this case becomes
\begin{equation}
   ds^2 = \mbox{Im}(\tau)\left|\frac{\eta^2(\tau)}
   {\sqrt{2}\Lambda^{2/3}}
   \prod_{i=1}^{2}
   (z-z_i)^{-1/12}dz\right|^2 \; ,
\end{equation}
where
\be
  z_1 = 8\Lambda^2 \quad , \quad z_2 = -8\Lambda^2
\ee
are the positions of the $(0,1)$ and $(2,1)$ 7-branes
respectively (see fig.~2).
The above metric
is related to the Seiberg-Witten solution as
\be
 T_{(p,q)}ds = |pda + qda_D| \; ,
\ee
which implies that the geodesic equation
for a $(p,q)$ string is given by
\be
  p\frac{da}{dt} + q\frac{da_{D}}{dt}
   = (p + q\tau)\frac{da}{dt} = c \; ,
\label{geodesic}
\ee
where $c$ is a constant. For topologically trivial
paths, {\it i.e.} paths which do not go around
7-branes, the solution is simply given by
\begin{equation}
 pa(z(t)) + qa_{D}(z(t)) = c(t-1) \; ,
\end{equation}
where $z(0)$ is the position of the D3-brane,
and $z(1)$ is the position of the $(p,q)$ 7-brane.
The geodesic is then gotten
by solving for $z(t)$.
The mass can now easily be calculated, and is
given by
\be
 m_{(p,q)} = |c| = |pa(z(0)) + qa_{D}(z(0))| \; ,
\ee
in agreement with the mass formula for BPS states
in Seiberg-Witten theory (\ref{swmass}). 
The only
states which correspond to topologically trivial
geodesics are the $(0,1)$, $(2,1)$ hypermultiplets
(assuming $N_f=0$), as these are the charges 
of the 7-branes (fig.~3a).

The hypermultiplet states with charges $(2n,1)$
($n>1$) would have to 
correspond to
strings along topologically non-trivial paths,
in order to pick up the appropriate monodromies
(fig.~3b,c).
One could solve eq.~(\ref{geodesic}) for fixed
values of $p$ and $q$, if we allow
$a$ and $a_D$ to be multivalued functions, by 
continuing them along the path. We shall denote
these continuations as $\tilde{a}$ and 
$\tilde{a}_D$. The solution is given by
\begin{equation}
  p\tilde{a}(z(t)) + q\tilde{a}_{D}(z(t)) = c(t-1) \; .
\end{equation}
Since $\tilde{a}$ and $\tilde{a}_D$ are not
single valued however, it is no longer obvious that
there exists a solution for $z(t)$, {\it i.e.}
a geodesic.

The W-boson, which carries the charges
$(2,0)$, would have to correspond
to a $(1,0)$ string that begins and ends on the D3-brane.
The orientations of the two ends must be the same,
and therefore the string must undergo a monodromy
which reverses its orientation. This is possible
if the string winds once around the $(0,1)$
and $(2,1)$ 7-branes (fig.~3d), 
for which the monodromy is
given by
\be
 M = \left(
  \begin{array}{rr}
    -1 & 4 \\
     0 & -1
  \end{array}
  \right) \; .
\label{monodromy}
\ee  
The solution satisfying this requirement is given by
\be
 2\tilde{a}(z(t)) = c(t-1/2) \; ,
\label{wgeodesic}
\ee
and it is again not obvious that one can solve for 
$z(t)$.
\begin{figure}[htb]
\epsfysize=4.3cm
\centerline{\epsffile{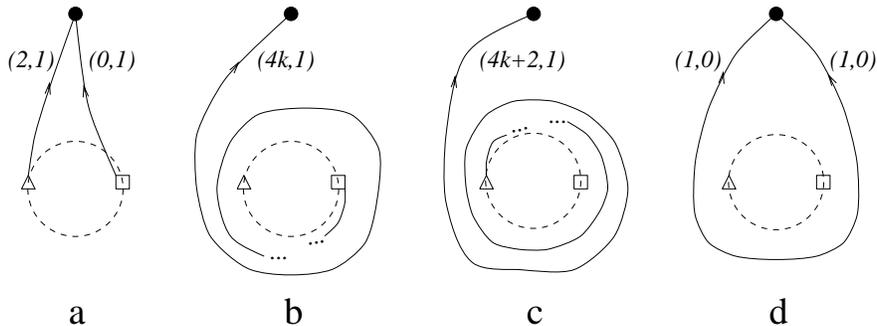}}
\caption{Paths corresponding to possible geodesics
for (a) the $(0,1)$ and $(2,1)$ hypermultiplets,
(b) hypermultiplets with charge $(4k,1)$ $(k>0)$,
(c) hypermultiplets with charge $(4k+2,1)$, and
(d) the W-boson.}
\end{figure}

In \cite{ansar} it was shown that there
are no topologically non-trivial geodesics
with base points in the strong coupling region
$\M_-$. This is consistent with the conjecture 
that the BPS spectrum in $\M_-$ consists of only the
$(0,1)$ and $(2,1)$ states, which correspond
to topologically trivial geodesics.
On the other hand, it is not at all clear that
non-trivial geodesics corresponding to the 
additional BPS states exist even in the weak
coupling region $\M_+$.
In fact one can argue that a geodesic corresponding
to the W-boson does not exist.
The equation for such a geodesic (\ref{wgeodesic})
implies that $\tilde{a}$ vanishes somewhere along
the geodesic (at $t=1/2$). For that to be the case,
the geodesic must pass through a 7-brane, whose 
$(p,q)$ charges agree with those of the string.
However, since the only 7-branes present carry
charges $(0,1)$ and $(2,1)$, and the charges of
the string vary from $(1,0)$ to $(1,1)$ and
back to $(1,0)$ along the path, this is impossible.
In section 4 we shall give another argument
against the existence of such a geodesic.

In the following we shall offer an alternative 
picture for the additional BPS states.
We shall argue that
these states are given by multi-pronged
strings connecting the D3-brane to the two 7-branes.
This picture has the advantage that it gives a clear 
physical explanation for the decay process. When
the prongs ending on the D3-brane degenerate, the
two other sets of prongs are free to separate along the
D3-brane. As we will show, this happens precisely
when the D3-brane is on the curve of marginal stability.

\section{Multi-pronged strings in 7-brane backgrounds}
\setcounter{equation}{0}

\noindent Multi-pronged strings are generalizations of 
ordinary $(p,q)$
strings in type IIB string theory, which possess more 
than two ends.
The simplest such object is the three-pronged string, 
consisting of
three strings with charges $(p_i,q_i)$ coming together 
at a junction
point, such that $\sum p_i=\sum q_i = 0$
\cite{asy,schwarz}. Other 
multi-pronged strings,
also known as string networks or string webs, are 
formed by connecting
a number of three-pronged strings.

Supersymmetry restricts the geometry of 
string networks \cite{dmjunction,sennetwork}. 
In the absence of 7-branes
the strings must be straight, the network must be planar,
and the relative orientations of the strings
are fixed according to their $(p,q)$ charges.
If we fix the coordinates
on the plane so that $(1,0)$ strings are oriented 
in the real direction, the condition for supersymmetry
given in \cite{dmjunction,sennetwork} is that $(p,q)$ 
strings are 
oriented in the $(p+q\tau)$ direction. In fact
an equally supersymmetric network is obtained if 
$(p,q)$ strings are oriented instead in the 
$(p+q\overline{\tau})$ direction. This network 
is related to the previous one by reflection about 
the real axis.
More generally, if we denote by 
$\phi_i$ the direction in the complex 
plane of a
$(p_i,q_i)$ string in a string network, the condition 
for supersymmetry is either
\be
  e^{i\phi_i} = e^{i\phi} {p_i + q_i \tau\over |p_i + 
q_i \tau|} \qquad\mbox{for all $i$} \;,
\label{flat1}
\ee
or 
\be
  e^{i\phi_i} = e^{i\phi}{p_i + q_i \overline{\tau}
\over |p_i + q_i \tau|} 
     \qquad\mbox{for all $i$}\; ,
\label{flat2}
\ee
where $\phi$ is the direction of the $(1,0)$ string.
In particular, this means that when a number of 
strings, with
$\sum p_i = \sum q_i = 0$, meet at a junction the 
relative angles
are fixed by supersymmetry in such a way as to give 
a vanishing force \cite{schwarz,dmjunction,sennetwork}.

The amount of supersymmetry preserved by string networks
is one fourth of the original amount in type IIB string
theory, {\it i.e.} 8 supersymmetries. If additional
D-branes are present the amount of unbroken 
supersymmetry depends on the type of D-branes and 
their orientation relative to the string network.
It was shown in \cite{bergman} that if the D-branes
are transverse to the string network, supersymmetry
is only preserved for D3-branes and D7-branes.
In the former case one has 4 supersymmetries, 
and in the latter case 8 supersymmetries.
The latter case can be generalized to any combination
of parallel $(p,q)$ 7-branes, which alone would
preserve the same 16 supersymmetries as a D7-brane.
This includes the background discussed in the previous
section.
Furthermore, including D3-branes in such backgrounds
will reduce the unbroken supersymmetries to 4.

In a generic background of 7-branes 
strings are not straight\footnote{There are special 
7-brane backgrounds 
for which $\tau$ is constant, and strings are 
therefore straight \cite{dm,ghz}.}, 
but rather follow geodesics (\ref{geodesic})
which are particular to the $(p,q)$ charges of the 
string.
The orientation of a $(p_i,q_i)$ string along
its geodesic is given by
\be
  e^{i\phi_i(z)} = {dz/dt_i \over |dz/dt_i|} 
    = {|p_i + q_i\tau(z)|\over (p_i + q_i\tau(z))} \,
        {|da/dz|\over da/dz}  \, {c_i\over |c_i|} \; .
\label{direction}
\ee
For a network of strings we must apply the 
supersymmetry condition
(\ref{flat1}) or (\ref{flat2}) locally. This gives 
either
\be
  e^{i\phi_i(z)} = e^{i\phi(z)} {p_i + q_i \tau(z)
\over |p_i + q_i \tau(z)|} 
    \qquad\mbox{for all $i$} \;,
\label{curved1}
\ee
or 
\be
  e^{i\phi_i(z)} = e^{i\phi(z)}{p_i + q_i 
\overline{\tau(z)}\over |p_i + q_i \tau(z)|} 
     \qquad\mbox{for all $i$}\; ,
\label{curved2}
\ee
where $\phi(z)$ is the local orientation of the $(1,0)$ 
string.
From (\ref{direction}) we see that this is given by
\be
  e^{i\phi(z)} = {|da/dz|\over da/dz} e^{i\phi} \; ,
\ee
where $\phi$ is an arbitrary constant angle, which 
corresponds to the
(constant) orientation of the $(1,0)$ string in flat 
space.
Consequently we deduce that only (\ref{curved2}) is 
consistent
with (\ref{direction})\footnote{It is not too 
surprising that only one
of (\ref{curved1}) or (\ref{curved2}) is consistent 
with supersymmetry,
since generic 7-brane backgrounds involve 
non-trivial monodromies
which break the reflection symmetry.}, 
and that 
\be
  c_i/|c_i| = e^{i\phi} \qquad \mbox{for all $i$} \; ,
\label{susy}
\ee
in other words that the phases of all the $c_i$ are equal.

\section{W-bosons}
\setcounter{equation}{0}

\noindent Let us now return to the particular 7-brane
background of section~2. In the classical picture the 
W-boson corresponds to a pair of fundamental strings
beginning on the D3-brane and ``ending'' on the 
$\Omega$7-plane. Quantum corrections deform
this background into two separated 7-branes, with
charges $(0,1)$ and $(2,1)$. At the same time the W-boson
state must somehow be deformed as well.
One possibility is that it becomes an open string which
winds around the two 7-branes (fig.~4a).
Another possibility
is that it is deformed into a four-pronged string,
with external prongs $(1,0), (1,0), (0,1)$ and $(2,1)$, 
and an internal $(1,1)$ prong 
\footnote{We will drop overall minus signs in $(p,q)$,
remembering that the orientations have to be such as
to satisfy charge conservation at each junction}, 
such that
the first two external prongs end on the D3-brane,
and the last two end on the corresponding 7-branes
(fig.~4b).

These two distinct configurations are in fact connected
by the process of string creation \cite{bdg}. 
In the simplest case this phenomenon occurs when
two mutually transverse D-branes cross.
In particular when a D-string, {\it i.e.} a $(0,1)$ 
string,
crosses a D7-brane, {\it i.e.} a $(1,0)$ 7-brane,
a fundamental $(1,0)$ string is created between them.
Using $SL(2,\bbbz)$ this can be generalized to a
$(p,q)$ string crossing a $(p',q')$ 7-brane with
$|pq' - p'q| = 1$, in which case a $(p',q')$ string
is created. The phenomenon can further be generalized
using charge conservation to arbitrary values
of $|pq' - p'q|$, in which case $|pq' - p'q|$ $(p',q')$
strings are created.
\begin{figure}[htb]
\epsfysize=4.5cm
\centerline{\epsffile{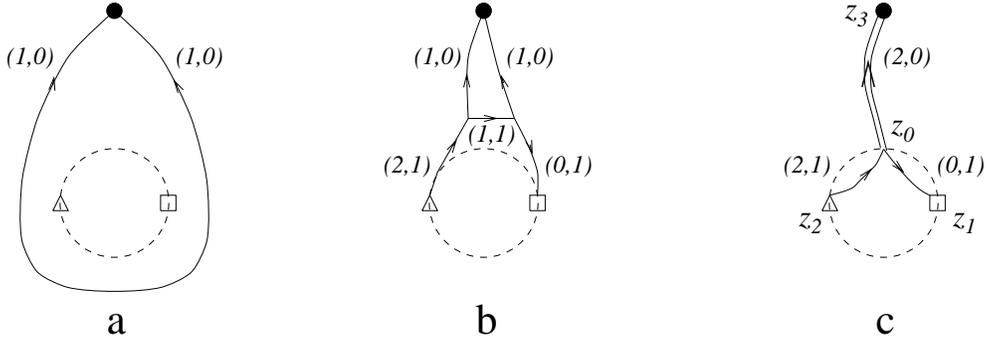}}
\caption{Possible quantum deformations of W-boson:
(a) A single string on a topologically non-trivial
path. (b) A four pronged string connecting the D3-brane
to the two 7-branes. (c) Four-pronged string with
degenerate intermediate prong. The last one is the only 
supersymmetric configuration.}
\end{figure}

Let us apply the supersymmetry conditions to the
above configurations. In both cases two ends
of a $(1,0)$ string end on the D3-brane with the 
same orientation. 
Supersymmetry requires that the two $(1,0)$
strings be parallel at the location
of the D3-brane. Since the geodesics for these 
strings are uniquely determined by their endpoint
and tangent at the endpoint, the two $(1,0)$ strings
must follow the same geodesic.
This condition cannot be satisfied for a single
string following a topologically non-trivial path
(fig.~4a), so such a path cannot correspond to a 
geodesic. 
On the other hand the four-pronged string
configuration (fig.~4b) will satisfy this condition
if the internal $(1,1)$ string shrinks to zero length,
and the two three-string junctions coincide (fig.~4c).

It should be noted that a similar conclusion which
rules out topologically nontrivial
closed geodesics was reached in \cite{ghz} without the use
of supersymmetry.

Let us denote the locations of the $(0,1)$ 7-brane, 
$(2,1)$ 7-brane, and D3-brane
by $z_1, z_2$, and $z_3$, respectively, and the 
location of the
two coincident junction points by $z_0$. 
Parameterize the geodesics for the 
$(0,1), (2,1)$, and $(1,0)$ strings by $t_1, t_2$ and
$t_3$, respectively, where $t_i=0$ corresponds
to the position of the junction, and $t_i=1$ to the 
position of the brane.
The three geodesics are then given by
\be
 \begin{array}{lrcl}
  P_1: & a_D(z(t_1)) &=& c_1 t_1 + a_D(z_0) \\
  P_2: & (-2a-a_D)(z(t_2)) &=& 
             c_2 t_2  -2a(z_0)-a_D(z_0) \\
  P_3: & a(z(t_3)) &=& c_3 t_3 + a(z_0) 
 \end{array}
\ee
where
\begin{eqnarray}
 c_1 &=& a_D(z_1) - a_D(z_0) = \mbox{} - a_D(z_0) 
     \nonumber \\
 c_2 &=& 2a(z_0)+a_D(z_0) - 2a(z_2) - a_D(z_2)
     =  2a(z_0)+a_D(z_0) \nonumber \\
 c_3 &=& a(z_3) - a(z_0) \; .
\label{c}
\end{eqnarray}
The second equality in the first two equations 
follows from the fact
that $pa+qa_D = 0$ at the location of a $(p,q)$ 7-brane.
The supersymmetry condition (\ref{susy}) is
\be
  {c_1/|c_1|} = {c_2/|c_2|} = {c_3/|c_3|} \; .
\label{susy2}
\ee
Using (\ref{c}), the first equality implies that
\be
 \mbox{Im}{a_D(z_0)\over a(z_0)} = 0 \qquad , \qquad
    {a_D(z_0)\over a(z_0)} > -2 \; ,
\label{marginal}
\ee
and the second equality implies that
\be 
  \mbox{Im}{a(z_3)\over a(z_0)} = 0 \qquad , \qquad
   {a(z_3)\over a(z_0)} > 1 \; .
\label{constphase}
\ee
The first condition requires $z_0$ to be 
on the curve of marginal
stability.
The second condition can be viewed as a condition on 
$z_3$,
{\it i.e.} the position of the D3-brane; namely that 
$z_3$ and $z_0$ be located
on a curve of constant phase for $a(z)$, and that 
$|a(z_3)|>|a(z_0)|$.
This is satisfied only if $z_3\in\M_+$, {\it i.e.} 
outside the curve 
of marginal stability (see Appendix~B for proof).

It follows from the behavior of $a(z)$ near $z=\pm 1$
(Appendix~A),
that its phase varies on the upper half plane part of
$\C$ from $0$ at $z=1$ to $\pi/2$ at $z=-1$. It also
follows from the asymptotic behavior for large $|z|$
that the phase varies on the upper half plane from $0$ at
$z\rightarrow +\infty$ to $\pi/2$ at
$z\rightarrow -\infty$. This implies that the
constant phase curves which cross the upper half
plane part of $\C$ foliate the upper half plane.
A similar argument holds for the lower half plane.

We conclude that the state corresponding to the
four-pronged string is BPS saturated if, and only if, 
the 
D3-brane is in the region $\M_+$.
When the D3-brane is exactly on the curve of marginal
stability, 
{\it i.e.} when $z_3=z_0$, the $(1,0)$ prongs degenerate, 
and the remaining
$(0,1)$ and $(2,1)$ strings can separate along the 
D3-brane (see fig.~5).
\begin{figure}[htb]
\epsfysize=4.0cm
\centerline{\epsffile{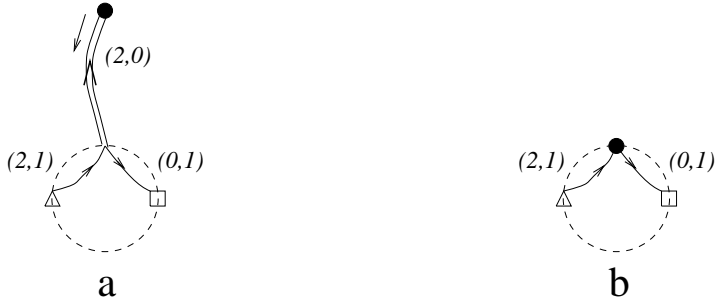}}
\caption{Decay of W-boson on curve of marginal stability.
When the D3-brane is on this curve, the two $(1,0)$ prongs
degenerate, leaving only the $(0,1)$ and
$(2,1)$ strings between the D3-brane and the respective
7-brane, which can now separate along the D3-brane.}
\end{figure}

The total mass is given by the sum of the four individual 
prong masses
\be 
 M = |c_1| + |c_2| + 2|c_3| \; .
\ee
The supersymmetry condition (\ref{susy2}) then implies 
that
\be 
 M = |c_1 + c_2 + 2c_3| = 2|a(z_3)| \; ,
\ee
which is precisely the (BPS) mass of the W-boson.

\section{Hypermultiplets}
\setcounter{equation}{0}

\noindent In addition to the W-bosons and the two
hypermultiplets carrying charges $(0,1)$ and $(2,1)$,
the BPS spectrum in the region $\M_+$ includes
hypermultiplets carrying charges $(2n,1)$, with
$n>1$. These are known to exist as solitons in
the weakly coupled field theory, and can in fact be
obtained from the $(0,1)$ and $(2,1)$ states by
performing monodromy transformations using
(\ref{monodromy}).

A single string configuration for these states
is shown in figs.~6a,b, and the corresponding 
topologically
trivial multi-pronged string configuration is
shown in fig.~6c. As before, the two configurations
are connected via string creation. 
Unlike the W-boson case however, it is not clear
whether supersymmetry forbids the single string
geodesic. On the other hand, we can still impose
the supersymmetry conditions on the multi-pronged
string. The geodesics for the three sets of prongs
are characterized by the complex numbers $c_i$,
which are now given by
\begin{eqnarray}
 c_1 &=& (1-n)a_D(z_0) \nonumber \\
 c_2 &=& n(2a(z_0) + a_D(z_0)) \nonumber \\
 c_3 &=& 2na(z_3) + a_D(z_3) - 2na(z_0) - a_D(z_0) \; .
\end{eqnarray}
Equating the phases of $c_1$ and $c_2$ gives
(\ref{marginal}) again, which restricts $z_0$
to lie on the curve of marginal stability.
Equating the phases of $c_2$ and $c_3$ then gives
\be 
  \mbox{Im}{2na(z_3)+a_D(z_3)\over 
2na(z_0)+a_D(z_0)} = 0 \qquad , \qquad
   {2na(z_3)+a_D(z_3)\over 2na(z_0)+a_D(z_0)} 
> 1 \; ,
\ee
which means that $z_3$ and $z_0$ must lie on the 
same curve of constant
phase for the function $f(z) = 2na(z) + a_D(z)$, 
and that $|f(z_3)| > |f(z_0)|$. As before,
this condition is satisfied only if $z_3 \in \M_+$
(see Appendix~B).
\begin{figure}[htb]
\epsfysize=4.5cm
\centerline{\epsffile{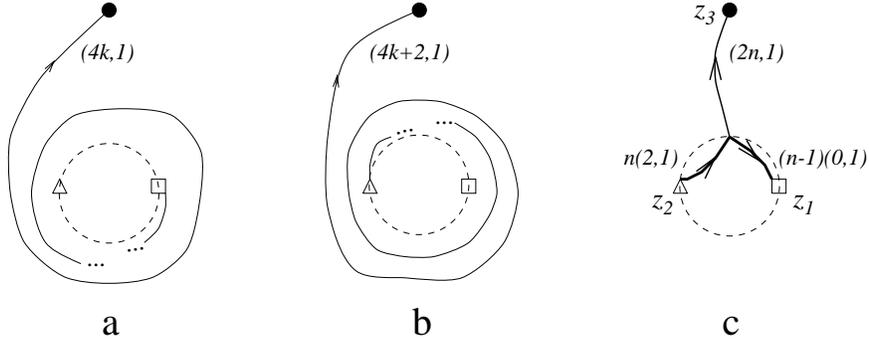}}
\caption{Possible configurations for the massive
hypermultiplets carrying charges $(2n,1)$.
These states could correspond to a single string
on a topologically non-trivial path, (a) in the
case $n=2k$, and (b) in the case $n=2k+1$,
or to a multi-pronged string (c).}
\end{figure}

As in the W-boson case, it follows that the multi-pronged
string in fig.~6c is BPS 
only if the D3-brane is in $\M_+$, 
and decays to $n$ $(2,1)$ strings
and $(n-1)$ $(0,1)$ strings on the curve of marginal 
stability. 
The mass of this state is given by
\be 
 M = |c_1| + |c_2| + |c_3| \; ,
\ee
which, due the equality of the phases of 
$c_1, c_2$ and $c_3$, becomes
\be 
 M = |c_1 + c_2 + c_3| = |na(z_3) + a_D(z_3)| \; ,
\ee
in precise agreement with the Seiberg-Witten result
(\ref{swmass}).

There is one subtlety we should point out.
Unlike the W-boson case, the constant phase curves
of the function $f(z)=2na(z)+a_D(z)$ which cross the 
upper half plane part of $\C$ do not foliate the upper
half plane. This can be seen from the behavior of
the phase of $f(z)$ on $\C$ and at infinity, which follows
from the asymptotic behaviors of $a$ and $a_D$ 
(Appendix~A). Specifically, the phase of $f(z)$
varies on the
upper half plane part of $\C$ from $0$
at $z=1$ to $\pi/2$ at $z=-1$. On the other hand,
at infinity 
it varies on the upper half plane
from $\pi/2$ at $z\rightarrow +\infty$
to $\pi$ at $z\rightarrow -\infty$,
and on the lower half plane from $0$ at
$z\rightarrow -\infty$ to $\pi/2$ at
$z\rightarrow +\infty$.
Consequently, the multi-pronged string state cannot
be BPS everywhere in $\M_+$, and we expect to see a 
transition to a single geodesic string state, of
the kind studied in \cite{ghz}.
This issue is currently under study \cite{bf2}

Let us consider more generally a state carrying
charges $(2n,m)$. For $|m|>1$ there does not exist 
a single string representation of this state.
One could however construct a suitable multi-pronged
string, consisting of 
a $(2n,m)$ prong \footnote{If $2n$ and $m$
are not relatively prime this would actually be
$k$ prongs, where $k$ is the largest common divisor.},
$n$ $(2,1)$
prongs, and $(n-m)$ $(0,1)$ prongs.
The geodesic constants are
given by
\begin{eqnarray}
 c_1 &=& (m-n)a_D(z_0) \nonumber \\
 c_2 &=& n(2a(z_0) + a_D(z_0)) \nonumber \\
 c_3 &=& 2na(z_3) + ma_D(z_3) - 2na(z_0) - ma_D(z_0) \; ,
\end{eqnarray}
for the $(0,1)$, $(2,1)$, and $(2n,m)$ prongs, 
respectively. The supersymmetry conditions are now
\be
  \mbox{Im}{a_D(z_0)\over a(z_0)} = 0 \quad , \quad
    \left\{
    \begin{array}{cc}
       a_D(z_0)/a(z_0) > -2 & \mbox{if } |m|<|n|\\
       a_D(z_0)/a(z_0) < -2 & \mbox{if } |m|>|n|
    \end{array}
    \right. \; ,
\label{marginal2}
\ee
and
\be 
  \mbox{Im}{2na(z_3)+ma_D(z_3)\over 
2na(z_0)+ma_D(z_0)} = 0 \qquad , \qquad
   {2na(z_3)+ma_D(z_3)\over 2na(z_0)+ma_D(z_0)} 
> 1 \; .
\ee
The first condition means that $z_0$ is again on the 
curve of marginal stability. However,
since $|a_D(z_0)/a(z_0)|<2$ on this curve,
a solution exists only if $|m|<|n|$.
This is consistent with the fact that from the
field theory point of view, states with $|m|>|n|$
would have to become massless somewhere on the 
curve of marginal stability, even though there
are no singularities on this curve, other than
the two corresponding to $(0,1)$ and $(2,1)$
\cite{bilalf}.
The second condition is satisfied, as before, 
only if $z_3 \in \M_+$ (Appendix~B).

It thus appears that BPS multi-pronged strings exist
for all states with charges $(2n,m)$, under the
condition that $|m|<|n|$ when $n>1$. This includes
in particular states with $|m|>1$, which do not
exist in the field theory. As argued in \cite{bilalf},
these states are related by the monodromy
at infinity (\ref{monodromy}) to states with
$|m|>|n|$. This argument is somewhat unsatisfactory
from the string theory point of view, as one would
like an argument based solely on the consistency
of a given string configuration. We do not have such 
an argument at this time.

\section{Conclusions}

\noindent We have demonstrated that the entire
BPS spectrum of Seiberg-Witten theory,
at strong coupling as well as at weak coupling,
can be accounted for in the D3-brane probe picture,
either as open strings on topologically trivial
geodesics between the D3-brane and a single
7-brane, or as multi-pronged strings connecting
the D3-brane to both 7-branes. The former correspond
to the hypermultiplets carrying charges $(0,1)$
and $(2,1)$, and exist everywhere in the moduli
space. The latter correspond to the W-bosons and
hypermultiplets carrying charges $(2n,1)$ ($n>1$),
and exist only in the region outside the curve 
of marginal stability. Their decay on this curve
is described by a simple string theory process
in which strings between the D3-brane and the 7-branes
separate along the D3-brane.

For the case of the W-boson we were able to show that
the multi-pronged string is the unique BPS 
representative. There are no topologically non-trivial
geodesics beginning and ending on the D3-brane
(see also \cite{ghz}).
An analogous argument for the massive hypermultiplets
is needed, since one clearly expects the BPS states
to be unique.

An equally pressing issue is the apparent existence
of BPS multi-pronged strings corresponding to states
which do not exist in the field theory, namely
hypermultiplets with charges $(2n,m)$, 
such that $|n|>|m|>1$.
One would like to rule
out these states at the string theory level by
some consistency argument, like the ``s-rule''
for example \cite{hw}.

\section*{Acknowledgments}

We would like to thank O. Aharony, A. Hanany,
T. Hauer and B. Zwiebach for useful discussions.
This work is supported in part by the NSF under grant 
PHY-92-18167.

\appendix
\section*{Appendix A}
\renewcommand{\theequation}{A.\arabic{equation}}
\setcounter{equation}{0}

\noindent In this appendix we summarize the asymptotic
behaviors of $a(z)$ and $a_D(z)$ near the singularities
$z=\pm 1, \infty$, as given in \cite{bilalf}.
The functions $a(z)$ and $a_D(z)$ can be expressed
in terms of hypergeometric functions as
\begin{eqnarray}
 a(z)& =& \left({z+1\over 2}\right)^{1/2}
     F\left(-\half,\half,1;{2\over z+1}\right)
    \nonumber\\
 a_D(z)&=&  i{z-1\over 2}
    F\left(\half,\half,2;{1-z\over 2}\right)\; .
\end{eqnarray}
Near the point at infinity, the asymptotic behaviors
are given by
\begin{eqnarray}
 a(z)&\sim& \sqrt{z/2}\nonumber \\
 a_D(z) &\sim& {i\over\pi}\sqrt{2z}
    \Big[\ln{z} + 3\ln{2} -2\Big] \; .
\label{zinf}
\end{eqnarray}
Near the branch point at $z=-1$:
\begin{eqnarray}
 a(z) &\sim& {i\over 2\pi}\left[
  \epsilon{z+1\over 2}\ln{z+1\over 2} 
  + {z+1\over 2}\Big(-i\pi-\epsilon(1+4\ln{2})\Big)
  + 4\epsilon\right] \nonumber \\
 a_D(z) &\sim& {i\over\pi}\left[
  - {z+1\over 2}\ln{z+1\over 2}
  + {z+1\over 2}(1+4\ln{2})
  - 4\right] \; ,
\label{z=-1}
\end{eqnarray}
and near the branch point at $z=1$:
\begin{eqnarray}
 a(z) &\sim& {2\over\pi} 
  - {1\over 2\pi}{z-1\over 2}\left[
    \ln{z-1\over 2} + 1 - 4\ln{2} \right] \nonumber\\
 a_D(z) &\sim& i{z-1\over 2} \; .
\label{z=1}
\end{eqnarray}

\section*{Appendix B}
\renewcommand{\theequation}{B.\arabic{equation}}
\setcounter{equation}{0}

\noindent In this appendix we would like to prove the
following theorem, relevant to sections 4 and 5:
\begin{thm}
  Let $f(z)=pa(z) + qa_D(z)$, and
  let $z_0$ be a point on the curve of marginal 
  stability $\C$, defined by
  $\mbox{Im}(a_D/a)=0$. Then for any point $z\in \M_+$,
  {\it i.e.} outside $\C$,
  such that $\mbox{Im}(f(z)/f(z_0)) = 0$, the 
  magnitude of $f$ satisfies
  $|f(z)|>|f(z_0)|$.
\end{thm}
The condition $\mbox{Im}(f(z)/f(z_0)) = 0$ means that
the points $z$ and $z_0$ lie on the same curve of 
constant
phase for the function $f(z)$.
We shall prove the above theorem in two steps. 
First we will show that 
$|f(z)|$ is monotonic along curves of
constant phase. Then we will show
that $|f(z)|$ is asymptotically increasing 
as we move away from $\C$ along curves of constant 
phase. The result $|f(z)|>|f(z_0)|$ then follows.

\vspace{5pt}
\noindent\underline{Step 1:}
Curves of constant phase are given by
\be
 {d\over dt}\mbox{Im}\Big(\ln f(z)\Big) = 
  {1\over 2}|f|^{-2}\Big(
 \overline{f}\dot{f} - f\dot{\overline{f}}\Big)
   = 0 \; .
\ee
This in turn implies that
\be
 {d\over dt}|f|^2 = 
  \overline{f}\dot{f} + f\dot{\overline{f}}
  = 2 \overline{f}\dot{f} \; ,
\ee
So 
\be
  {d\over dt}|f|^2 = 0 \quad \Longleftrightarrow
  \quad f=0 \quad \mbox{or} \quad \dot{f}=0 \; .
\ee
However $f$ can only vanish for $(p,q)=(0,1)$ or $(2,1)$,
and this happens at the locations of the 7-branes. 
As we are interested in other values of $(p,q)$,
we conclude that $|f|$ can have an extremum only
if $\dot{f}$ vanishes somewhere. This would imply
\be
 p\dot{a} + q\dot{a}_D = 0 \; ,
\ee
and therefore that 
\be
 \tau\equiv {da_D\over da} = -{p\over q} \in\bbbr \; .
\ee
On the other hand positivity of the metric on moduli 
space
requires $\mbox{Im}(\tau) > 0$, so $\dot{f}$ cannot
vanish anywhere. $|f(z)|$ is therefore monotonic
along curves of constant phase.

\noindent\underline{Step 2:}
From the asymptotic behaviors of $a$ and $a_D$
as $|z|\rightarrow\infty$ (\ref{zinf}) 
we see that
\be
 f(z)\sim iq\sqrt{r}\ln{r}e^{i\theta/2}\qquad
 \mbox{as} \quad r\rightarrow\infty \; ,
\ee
where $z=re^{i\theta}$. 
It follows that in the limit
$r\rightarrow\infty$ 
the constant phase curves are asymptotic to radial
lines, and that 
\be
 {d|f|\over dr} > 0 \; .
\ee
$|f(z)|$ is therefore asymptotically increasing 
along the curves of constant phase.

\vspace{5pt}
\hfill {\bf QED}

\end{document}